\documentclass[aps,pre,twocolumn,superscriptaddress,floatfix]{revtex4-1}
\usepackage[utf8]{inputenc}
\usepackage{graphicx}
\usepackage{amsmath}
\usepackage{amsfonts}
\usepackage{amssymb}
\usepackage{subcaption} 
\usepackage{epsfig}
\usepackage{epstopdf}
\usepackage{pstool}
\usepackage{psfrag}
\usepackage[pagewise,displaymath,mathlines]{lineno}
\usepackage{tikz}
\usepackage{xcolor}
\usepackage{mathtools}

\usepackage{pgfplots}
\DeclareUnicodeCharacter{2212}{−}
\usepgfplotslibrary{groupplots,dateplot}
\usetikzlibrary{patterns,shapes.arrows}
\pgfplotsset{compat=newest}

\usepackage{array}
\usepackage{multirow}
\usepackage{setspace}

\usepackage{caption}
\usepackage{epstopdf}
\usepackage[normalem]{ulem}
\usepackage{float}
\usepackage{ifsym}

\def\pd{\partial}
\def\reals{{\mathbb R}}

\newcommand {\pdTwo}[2]{\frac{\partial^{2} #1}{\partial #2^{2}}}

\newcommand {\upd}{\mathrm{d}}

\newcommand {\phiend}{\phi_{\mathrm{end}}}
\renewcommand{\i}{\mathrm{i}}

\newcommand {\taue}{\tau^{(\text{s})}}
\newcommand {\tauo}{\tau^{(\text{a})}}
\newcommand {\tauu}{\tau^{(1)}}
\newcommand {\taud}{\tau^{(2)}}

\newcommand {\tL}{\tilde{L}}

\renewcommand{\varphi}{\chi}

\def\tb{\textcolor{blue}}

\newcommand{\nordita}{Nordita, Royal Institute of Technology and Stockholm University, Stockholm 10691, Sweden}

\newcommand{\be}{\begin{equation}}
\newcommand{\ee}{\end{equation}}
\newcommand{\ba}{\begin{align}}
\newcommand{\nn}{\nonumber}

\begin{document}

\preprint{APS/123-QED}

\title{Wrinkling composite sheets}

 \author{Marc Su\~n\'e}
\email{marc.sune.simon@su.se}
\affiliation{\nordita}

 \author{Crist\'obal Arratia}
\email{cristobal.arratia@su.se}
\affiliation{\nordita}

 \author{A. F. Bonfils}
\email{anthony.bonfils@su.se}
\affiliation{\nordita}

\author{Dominic Vella}
\email{dominic.vella@maths.ox.ac.uk}
\affiliation{Mathematical Institute, University of Oxford, Woodstock Rd, Oxford, UK, OX2 6GG}

\author{J.S. Wettlaufer}
\email{john.wettlaufer@yale.edu, jw@fysik.su.se}
\affiliation{Yale University, New Haven, Connecticut 06520, USA}
\affiliation{\nordita}

\begin{abstract}
	We examine the buckling shape and critical compression of confined inhomogeneous composite sheets lying on a liquid foundation. The buckling modes are controlled by the bending stiffness of the sheet, the density of the substrate, and the size and the spatially dependent elastic coefficients of the sheet. We solve the beam equation describing the mechanical equilibrium of a sheet when its bending stiffness varies parallel to the direction of confinement. The case of a homogeneous bending stiffness exhibits a degeneracy of wrinkled states for certain lengths of the confined sheet; we explain this degeneracy using an asymptotic analysis valid for long sheets, and show that it corresponds to the switching of the sheet between symmetric and antisymmetric buckling modes. This degeneracy disappears for spatially dependent elastic coefficients. Medium length sheets buckle similarly to their homogeneous counterparts, whereas the wrinkled states in large length sheets concentrate the bending energy towards the soft regions of the sheet.
\end{abstract}

\maketitle

\date{\today}

\section{Introduction \label{sec:Introduction}}

From surfactant monolayers to tectonic plates, the deformation of elastic sheets underlies a vast landscape of problems in science and engineering. Not only do their patterns evoke a great aesthetic appeal, but they underlie both function and form in the world around us \cite{Vella:2007, Vella:2008, HudlestonTreagus10, Mansfield}.

The buckling of a homogeneous sheet on top of a supporting foundation has been the object of soft matter studies for decades~\cite[see e.g.,][and Refs.~therein]{MilnerJoanny89, CerdaMahadevan03, VellaAussillous04}. In a free--standing confined sheet (Euler's elastica), the  compressive load (which we refer to as ``compression" throughout the paper) at which buckling occurs, and the scale of this buckling, are set by the length $L^*$ of the confined sheet. However, in the presence of a supporting foundation, an intrinsic length scale $\ell^*$ is set by the mismatch between the elasticity of the sheet and that of the foundation. When a resting sheet with clamped ends is confined longitudinally, a uniaxial compression builds up and wrinkles of the surface emerge \cite{BourdieuDaillant94}.
Such wrinkle patterns have been observed in, among many other settings, thin polymer sheets resting on the surface of water~\cite{PocivavsekDellsy08}, bacterial biofilms~\cite{TrejoDouarche13}, granular rafts~\cite{Jambon-PuilletJosserand17}, the mineral veins of rocks~\cite{HudlestonTreagus10}, strained epitaxial films \cite{Spencer}, and in glaciology \cite{Kerr1976,MacAyeal}.
From a theoretical standpoint, wrinkles of confined elastic sheets are predicted by a simple set of geometric rules~\cite{TobascoTimounay22}. 
The theory for supported sheets also maps onto a model to describe the elasticity of an unsupported epithelial monolayer~\cite{AndrensekZiherl23}.

In many settings the deforming material is treated as a homogeneous solid, despite a potentially important intrinsic composite structure. Indeed, whether or not the composite structure can be treated as effectively homogeneous requires appropriate quantitative analysis. When the host material is stiff, such as ice, and the inclusions are soft, the theory of Eshelby~\cite{Eshelby57} shows that the effective Young's modulus, and hence the bending stiffness is reduced relative to the case of the stiff host material alone.  When the host solid is soft and the inclusions are uniformly distributed~\cite{StyleBoltyanskiy15}, two possibilities exist---composite softening or stiffening---depending on the size of  inclusions relative to the elastocapillary length, $l^*\equiv\gamma^*/E_{0}^*$, where $\gamma^*$ is the surface tension of the inclusion-host interface and $E_{0}^*$ is the Young's modulus of the host material. However, by controlling the distribution of inclusions we can, for example, prepare a gradient of bending stiffness, and thereby manipulate the failure properties of such elastic composites.

Here we examine the deformation of composite elastic sheets floating on a liquid foundation. We treat the case with a gradient of bending stiffness parallel to the direction of confinement. This gradient is envisaged to be created by controlling the distribution of liquid inclusions.

First, we present key results for the critical compressive load and wrinkle patterns in homogeneous floating sheets.
The interplay between the intrinsic length scale, $\ell^*$, and the undeformed size of the confined sheet, $L^*$, determines the wrinkled state that has the smallest compression. The observed wrinkles are then either symmetric or antisymmetric about the centre of the sheet.
Importantly, at certain sheet lengths the wrinkled states are degenerate: both symmetric and antisymmetric modes exist at the same compression.

While the homogeneous case provides important insights into the wrinkle pattern, our main interest is the study of inhomogeneous sheets. Taking advantage of the effective medium behaviour of composites~\cite{Eshelby57,StyleBoltyanskiy15,MancarellaStyle16}, we examine the wrinkles in composite sheets. 

We impose a spatial distribution of liquid inclusions that translates into spatially varying elastic moduli (Young's modulus and Poisson ratio) and hence a spatially varying bending stiffness. The inhomogeneous stiffness is responsible for breaking the symmetry of the wrinkle patterns and eliminating the degeneracies. For sheets large compared to the intrinsic length scale, this gradient shifts the position of the maximum amplitude of wrinkles. This displacement happens at the onset of wrinkling, and hence it does not require any non--linear behaviour. Inhomogeneous sheets are technologically appealing for they offer a new means to control fracturing via the spatial variation of the wrinkle amplitude.

\section{Mechanics of a floating composite sheet\label{sec:mech}}
We start with a presentation of the governing equations for a composite thin sheet floating on a liquid, including the effects of in--plane forces. For more detail the reader is referred to the book by Mansfield \cite{Mansfield}.

\subsection{Problem formulation\label{subsec:formulation}}

\begin{figure}
	\input{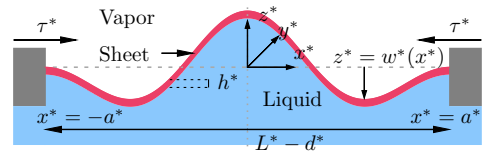}
	\caption{Schematic diagram showing a thin elastic sheet floating on a liquid with a lateral compressive force $\tau^*$ due to confinement.}
\label{mech:schema}
\end{figure}

We denote all dimensional quantities with a superscript  $(\cdot)^*$ and let $x^*,y^*,z^*$ be the Cartesian coordinates in the horizontal direction, into the page, and in the vertical direction, respectively.
We consider a sheet with thickness $h^*$ and longitudinal undeformed length $L^*$, such that $h^*\ll L^*$. The sheet (of density $\rho^*_{\text{s}}$) rests on a liquid of density $\rho^*$, and in the absence of all forces apart from gravity, it floats with its mid-line at a height $z^*= h^*(1/2-\rho^*_\text{s}/\rho^*)$ above the free surface.  We measure all vertical displacements relative to this equilibrium level~\footnote{We treat sheets floating on water with density $\rho^*\sim10^3\,\tfrac{\text{kg}}{\text{m}^3}$, and the standard value of the gravitational acceleration $g^*\approx 9.8\,\tfrac{\text{m}}{\text{s}^2}$.}, as in Fig.~\ref{mech:schema} where the origin of the vertical axis is set at the liquid level. The sheet is confined lengthwise by a distance $d^*$, and has clamped edges located at $x^*=\pm a^*$ in the longitudinal direction, with $a^*\equiv (L^*-d^*)/2$ .
However,  we shall impose boundary conditions at $x^*=\pm L^*/2$, which are the Lagrangian coordinates of the sheet's edges; this avoids confusion with their position at the onset of buckling, $x^*=\pm a^*$, and is correct within small deformation elasticity. A quantity of interest from our analysis is the value of $a^*$ at the onset of buckling for a given sheet length $L^*$.

The geometric incompatibility between the sheet's undeformed length and the confinement length yields a mechanical instability producing a vertical displacement $w^*(x^*,y^*)$. This displacement causes, and is influenced by, the restoring pressure $p^*=-\rho^*g^*w^*$ that the liquid foundation exerts on the sheet. An  in--plane compression $\tau^*$ (with dimensions of force per unit length) is calculated as an emergent property, rather than being imposed.

\begin{figure}
\centering
\includegraphics[width=\columnwidth]{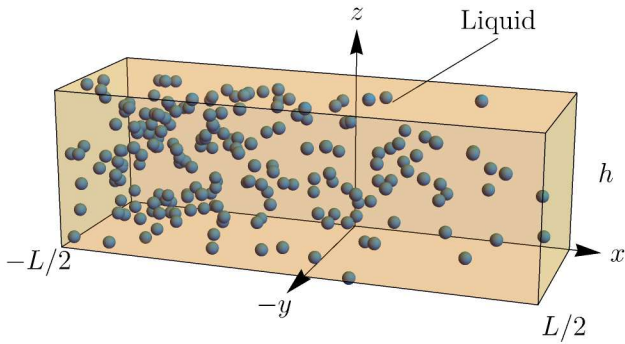}
        \caption{Strip of a composite sheet of length $L$, with liquid inclusions and the linear volume fraction profile of Eq.~\eqref{buckle:phi} in the $x$ direction, parallel to the confinement.
        }
\label{buckle:strip}
\end{figure}

The static equilibrium of a confined floating sheet is determined by the balance between elastic forces and the hydrostatic pressure. In this paper, we are concerned with the effects of a variable bending stiffness, or flexural rigidity, $B^*(x^*)$, which may be achieved using
 composite materials whose elastic properties are inhomogeneous. Such inhomogeneous sheets can be made by embedding one material in the other to control the resulting composite structure, as illustrated schematically in Fig.~\ref{buckle:strip}.
Examples include ionic liquid droplets in silicone~\cite{StyleBoltyanskiy15}, hydrogel particles inside elastomeric matrices~\cite{MoserFeng22}, 3D printing materials using silicone double networks~\cite{WallinSimonsen20} and fibre-silicone mixtures~\cite{MoLong22},
or hydrogel substrates with
a photo-sensitive cross-linker 
and a gradient photo-mask \cite{Tse2010}. Despite a varying composition, we neglect any variation of the density $\rho_{\text{s}}^*$ of the sheet.
This is justified for stiff composites in the dilute regime, and for soft composites, because the matrix and the inclusions have approximately the same density.
(A varying bending stiffness would also correspond to a single component material sheet with varying thickness, but we do not treat this case here because it would complicate the free-floating equilibrium.) 

In the absence of inclusions, namely for a homogeneous elastic sheet, the elastic coefficients (Young's modulus, $E_{0}^*$, and Poisson ratio, $\nu_{0}$; we denote the elastic constants of the homogeneous sheet with subscript ``0") are constant, and hence so too is the bending stiffness;  in particular, the baseline bending stiffness is
\begin{equation}
	B_{0}^*=\frac{E_{0}^*\,{h^*}^3}{12(1-\nu_{0}^2)}.\label{mech:bending_hom}
\end{equation}
The intrinsic length scale of the displacements, or wrinkles, 
\be
\ell^*\equiv \left(\frac{B_{0}^*}{\rho^* g^*}\right)^{1/4}\,,
\ee plays a central role in the problem.  Therefore, we use $\ell^*$ to rescale all lengths, while we rescale the compressive force $\tau^*$ with  $B_0^*/(\ell^*)^2$ . Dimensionless quantities are unstarred so that $x=x^*/\ell^*$, and $\tau=\tau^*/(B_0^*\rho^*g^*)^{1/2}$. This non-dimensionalization also introduces a typical scale for the size of the stresses induced by compression, which we define as
\begin{equation}
        {\cal S}\equiv\frac{E_{0}^*h^*}{(B_{0}^*\rho^* g^*)^{1/2}}\,,\label{mech:stretching}
\end{equation} 
which is also a measure of the relative ease of bending to stretching of the sheet, or the stretching--stiffness. Therefore, ${\cal S}$ is important in determining how much confinement is required to induce buckling, as we describe later. Because $B_{0}^*\sim E_{0}^* {h^*}^3$ so that ${\cal S}\sim [E_{0}^*/(\rho^* g^*h^*)]^{1/2}$, for thin sheets we expect that ${\cal S}\gg1$, and hence deformation can be accommodated more easily by bending than by stretching. For example, a 10 cm thick sheet of ice, for which $E_{0}^* \sim$ GPa and $\nu_{0} \approx 0.3$, has ${\cal S}\sim$ 1000 \cite{Vella:2008}. A soft material, such as PDMS, for which $E_{0}^* \sim$ 100 kPa and $\nu_{0} \approx 0.5$, has ${\cal S}\sim 100$ for $h^* $ on the order of cm. However, thin sheets do not always bend more easily than they stretch. Provided that ${\cal S}$ is finite, sheets may accommodate some of the imposed deformation by compressing (negative stretching), particularly when the confinement is small and the supported sheet resists buckling. We therefore refer to ${\cal S}$ as the `inextensibility' of the sheet. Before returning to this idea, we introduce buckling in two dimensions where the equations of mechanical equilibrium simplify.

\subsection{Two--dimensional buckling of a sheet\label{buckle}}

Our focus here is on two-dimensional deformations of the sheet under uniaxial compression. A simple argument (see \tb{Appendix \ref{App:GovEqn}}) shows that the in-plane stress $\tau$ is constant and that small out-of-plane displacements of the sheet satisfy
\begin{equation}
        \frac{\mathrm{d}^2}{\mathrm{d}x^2}\left[B(x) w_{,2x}\right]+\tau w_{,2x}+w=0,
\label{buckle:ode}
\end{equation}
which is Euler's linearized elastica equation \cite[see e.g. \S 20 of Ref.~][]{LandauLifshitz86} with a lateral load due to the hydrostatic pressure in the liquid foundation and varying elastic properties along the axis of confinement.
This is analogous to the small--deflection equilibrium of a compressed inhomogeneous column~\cite{SuneWettlaufer21}.
The boundaries of the thin sheet at $x=\pm L/2$ are clamped so that $w_{,x}(\pm L/2)=w(\pm L/2)=0$. 

The elastic response of a composite is characterized by the elastocapillary length $l^*$ discussed in the Introduction (\S\ref{sec:Introduction}). When inclusions of radius $R^*$ are smaller (larger) than $l^*$, the composite is difficult (easy) to deform because surface tension can maintain (cannot maintain) their sphericity. For example, for a typical liquid inclusion with $\gamma^*=O(0.01\text{N\,m}^{-1})$~\cite{StyleBoltyanskiy15}, and soft materials with a range of $E^*=O(1-100\text{kPa})$, $l^*$ ranges from $1$ to $100\mu \text{m}$. Thus, surface tension effects are important for micron--sized inclusions. A summary of different theoretical models that predict the bulk mechanical properties of a composite is given in \tb{Appendix \ref{AppA}}.

\section{Buckling of a homogeneous sheet\label{homogeneous}}

The force balance of an elastic sheet with homogeneous properties, namely $B(x)=1$ in Eq.~\eqref{buckle:ode}, is given by
\begin{equation}
        (w_0)_{,4x}+\tau_0 (w_0)_{,2x}+w_0=0,
\label{homogeneous:ode0}
\end{equation}
where we use subscript ``$0$" to denote the displacement in the homogeneous case. This is to be solved with boundary conditions corresponding to a clamped sheet: 
\begin{subequations}
\begin{equation}
	w_0(\pm L/2)=0,\quad \text{and} \quad	(w_0)_{,x}\Big|_{x=\pm L/2}=0.\tag{\theequation a-b}
\end{equation}
	\label{homogeneous:bc}
\end{subequations}

\subsection{Buckling profiles}
Equations~\eqref{homogeneous:ode0} and~\eqref{homogeneous:bc} define an eigenvalue problem whose detailed solution is given in \tb{Appendix \ref{AppB}}. As expected from the reflection symmetry about $x=0$, there are two distinct types of solutions with opposite parity: even functions of the form $w^{\text{(s)}}_0(x)=A_{\text{s}}\cos kx$, which correspond to symmetric buckling profiles, and the odd counterpart $w^{\text{(a)}}_0(x)=A_{\text{a}}\sin kx$, giving antisymmetric profiles. For both cases the wavenumber $k$ satisfies
\begin{equation}
k^4-\tau_0 k^2+1=0.
\label{homogeneous:quartic}
\end{equation}
Therefore, there are two possible wavenumbers, $k_\pm$, given by
\begin{equation}
k_\pm^2=\tfrac{1}{2}\bigl(\tau_0\pm\sqrt{\tau_0^2-4}\bigr).
\label{homogeneous:waveno}
\end{equation}
Importantly, the domain of $k_{\pm}(\tau_0)$ (such that $k_{\pm}(\tau_0)\in\reals$) is bounded from below, where $\tau_0=2$ gives the value where the two branches of the solutions meet, $k_+(\tau_0)=k_-(\tau_0)$. We show in \tb{Appendix \ref{App:RealWaveNo}} that there are no solutions of the eigenvalue problem (Eqs.~\ref{homogeneous:ode0},~\ref{homogeneous:bc}) for complex $k_\pm$.

In general, $w_0(x)$ will contain both of the wavenumbers given in Eq. \eqref{homogeneous:waveno}. The condition of zero vertical displacement at $x=\pm L/2$ is satisfied by the symmetric and antisymmetric combinations viz.,
\begin{subequations}
\begin{align}
	w_0^{(\text{s})}(x)&=A_{\text{s}}\left(\frac{\cos k_+x}{\cos k_+L/2}-\frac{\cos k_-x}{\cos k_-L/2}\right),\label{homogeneous:sole}\\
	\text{and}\quad w_0^{(\text{a})}(x)&=A_{\text{a}}\left(\frac{\sin k_+x}{\sin k_+L/2}-\frac{\sin k_-x}{\sin k_-L/2}\right),
\label{homogeneous:solo}
\end{align}
\label{homogeneous:sols}
\end{subequations}
\noindent where $A_{\text{s}},A_{\text{a}}$ are, yet to be determined, constants. The remaining boundary condition that $(w_0)_{,x}(\pm L/2)=0$ leads to the relations between $k_+$ and $k_-$;
\begin{subequations}
\begin{align}
	k_+\tan k_+\frac{L}{2}&=k_-\tan k_-\frac{L}{2},\label{homogeneous:disprele}\\
	\text{or}\qquad k_+\cot k_+\frac{L}{2}&=k_-\cot k_-\frac{L}{2},\label{homogeneous:disprelo}
\end{align}
\label{homogeneous:disprels}
\end{subequations}
which are associated with the symmetric~\eqref{homogeneous:sole} and antisymmetric~\eqref{homogeneous:solo} modes respectively. Since $k_\pm=k_\pm(\tau_0)$, the solutions of Eq. \eqref{homogeneous:disprele} determine the compression $\taue_0$ required to produce the corresponding symmetric profile $w_0^{(\text{s})}(x)$ from Eq. \eqref{homogeneous:sole}. Analogously, the roots $\tauo_0$ of equation~\eqref{homogeneous:disprelo} are associated with the odd functions $w_0^{(\text{a})}(x)$ from Eq.~\eqref{homogeneous:solo}. For a given value of $L$, each relation \eqref{homogeneous:disprele} and \eqref{homogeneous:disprelo} has an infinite number of solutions, the smallest of which is $\tau_0=2$. However, when $\tau_0=2$, $k_+=k_-$ and Eqs. \eqref{homogeneous:sols} give the trivial solution $w_0(x)\equiv0$. Therefore, each value of $\tau_0>2$ that solves either of Eqs. \eqref{homogeneous:disprels} corresponds to a different, and non-trivial mode of buckling in the sheet. We are concerned with determining only the lowest mode of buckling, which corresponds to the smallest value of $\tau_0>2$ that solves either Eq.~\eqref{homogeneous:disprele} or Eq.~\eqref{homogeneous:disprelo} for a given sheet length $L$.

\subsection{Asymptotic results for large sheets\label{sec:Asymptotics}}

\begin{figure}
\input{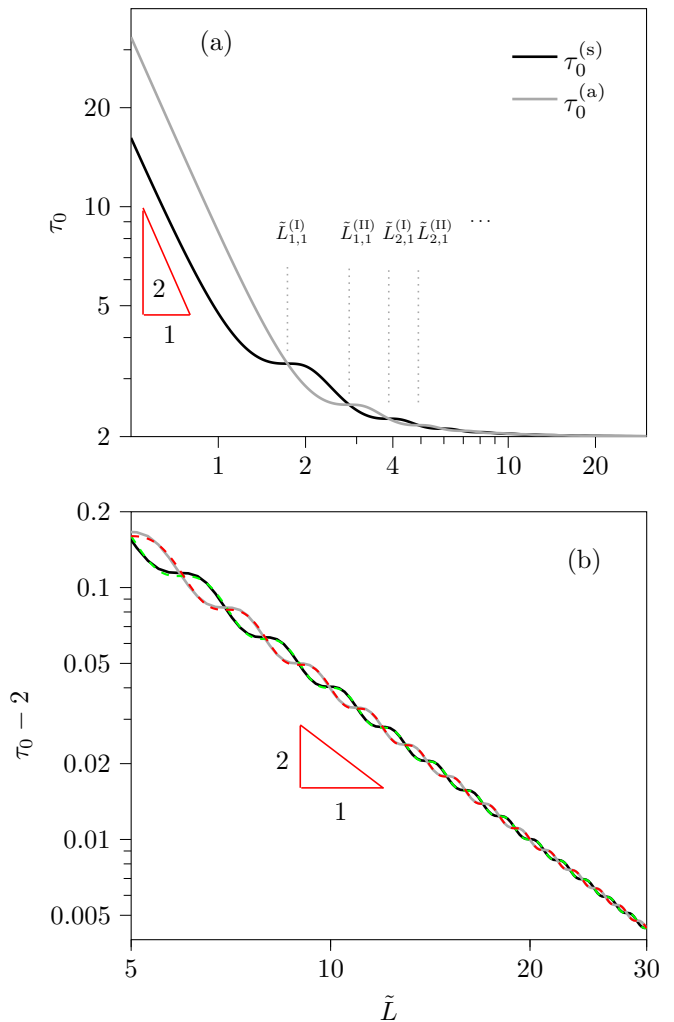}
	\caption{Minimum compressive force $\tau_0$ as a function of the size of the confined sheet $L$. (a) $\tau_0(L)$ for medium length sheets on a log--log plot. (b) Log--log plot of $\tau_0(L)-2$ for large  sheets showing the good agreement between numerical solutions of \eqref{homogeneous:disprele} (black solid) and \eqref{homogeneous:disprelo} (gray solid) and the asymptotic relationships \eqref{eqn:Tau0AsyEven} (green dashed) and \eqref{eqn:Tau0AsyOdd} (red dashed), respectively. 
	}
\label{finite-size:modes}
\end{figure}

Figure \ref{finite-size:modes} shows numerical results for the two smallest values of $\tau_0$ at the onset of instability as a function of the sheet length $L$. The quantity of principal interest is the smallest buckling load, $\tau_0>2$. These numerical results suggest that the critical compressive load at the onset of wrinkling corresponds to a symmetric or antisymmetric mode depending on the precise value of $L$. To understand this behaviour, we also note that the critical buckling load appears to approach $2$ from above as the sheet length $L\to\infty$. We use this as motivation to seek an asymptotic solution of Eqs. \eqref{homogeneous:disprele}--\eqref{homogeneous:disprelo} for $\epsilon=\tau_0-2\ll1$. This analysis (see \tb{Appendix \ref{App:Asymptotics}}) reveals that:
\begin{equation}
\tau_0^{(\text{s})}=2+\frac{4\pi^2}{L^2}\left(1+2\frac{\sin L}{L}\right)+O(1/L^4)
\label{eqn:Tau0AsyEven}
\end{equation} for symmetric modes, while
\begin{equation}
\tau_0^{(\text{a})}=2+\frac{4\pi^2}{L^2}\left(1-2\frac{\sin L}{L}\right)+O(1/L^4)
\label{eqn:Tau0AsyOdd}
\end{equation} for antisymmetric modes. These results go to higher order in $L^{-1}$ than the equivalent result by \citet{Rivetti2014}.

Note that, in either case, the result that $\tau_0\to2$ suggests that $k_{\pm}\to1$ and hence that the natural dimensionless wavelength is $\lambda=2\pi$. It is therefore often useful to measure the sheet length in terms of the number of half wavelengths and so we introduce 
\begin{equation}
\tL=L/\pi.
\label{homogeneous:atilde}
\end{equation} We use $L$ and $\tL$ interchangeably for convenience.

The asymptotic predictions \eqref{eqn:Tau0AsyEven} and \eqref{eqn:Tau0AsyOdd} show excellent agreement with our numerical solutions of \eqref{homogeneous:disprele}--\eqref{homogeneous:disprelo} for $\tL\gtrsim 8$ (see Fig.~\ref{finite-size:modes}~(b)). They also illustrate a $\tL^{-2}$ power-law (red triangles in both Fig.~\ref{finite-size:modes} (a) and (b)) and how the mode with smallest critical compression oscillates between the symmetric and antisymmetric modes as $\sin L$ oscillates. Asymptotically we have that the mode at onset should be antisymmetric when $\sin L>0$ (i.e.~when $2n\pi <L<(2n+1)\pi$), while it should be symmetric when $\sin L<0$, (i.e.~when $(2n-1)\pi<L<2n\pi$),  for integer $n$. A more detailed discussion of this feature, for general values of $L$, is given in \tb{Appendix \ref{App:Crossings}}.

The asymptotic results for the critical loads also allow us to give simpler expressions for the symmetric and antisymmetric mode shapes in the limit $L\gg1$ (see \tb{Appendix \ref{App:Asymptotics}}):
\begin{subequations}
\begin{align}
	w_0^{(\text{s})}(x)&=A_{\text{s}}\cos x \,\cos\left(\frac{x}{\tL}\right),\label{homogeneous:asympe}\\
	\text{and}\quad w_0^{(\text{a})}(x)&=A_{\text{a}}\sin x \,\cos\left(\frac{x}{\tL}\right).
\label{homogeneous:asympo}
\end{align}
\label{homogeneous:asymp}
\end{subequations}
These results show how the mode shapes consist of a sinusoid of (short) wavelength, equal to $2\pi$, whose amplitude is modulated by another sinusoid with a large wavelength equal to $2L$. This result also makes clear that the amplitude of the wrinkles varies spatially as a result of a beating phenomenon --- this spatial variation of wrinkle amplitude is similar to, but has a simpler underlying cause than, the spatial variation observed by Tovkach \emph{et al.} \cite{Tovkach2020}.

\begin{figure}
\input{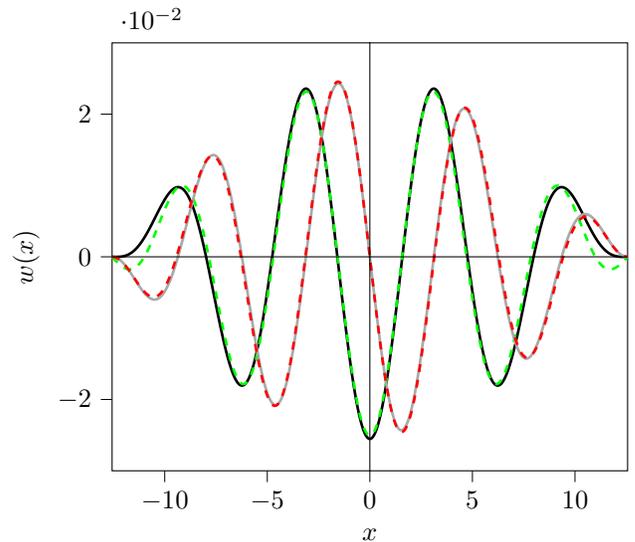}
	\caption{Mode shapes corresponding to the two smallest compressions for a sheet of size $\tL=8$. Results are computed from the exact solutions, Eqs.~\eqref{homogeneous:sols} plotted with solid lines, and the asymptotic expressions, Eqs.~\eqref{homogeneous:asymp} in dashed lines. The amplitudes are computed assuming the condition $(a_c-a)=10^{-3}\ll1$, which ensures that the sheet bends close to the onset of buckling, imposed in Eq.~\eqref{homogeneous:intrel}.}
\label{asymptotics:modes}
\end{figure}

Figure~\ref{asymptotics:modes} shows a comparison between the asymptotic expressions of \eqref{homogeneous:asymp} with the exact mode shapes predicted by Eqs.~\eqref{homogeneous:sols}, for a moderately large sheet ($\tL=8$). The agreement between the asymptotic predictions and the shapes given by Eqs.~\eqref{homogeneous:sole}--\eqref{homogeneous:solo} is good, and small deviations only become visible near the ends, which is more pronounced 
for the symmetric mode, shown in the green dashed line of Fig.~\ref{asymptotics:modes}.  Note that the boundary condition Eq.~(\ref{homogeneous:bc}b) is only satisfied at $O(L^{-1})$.

Our asymptotic results are related to those reported by~\citet{PocivavsekDellsy08}. Indeed, the linear combination of $A_{\text{s}}=\sin (L/2)$ in Eq.~\eqref{homogeneous:asympe} and $A_{\text{a}}=\cos (L/2)$ in Eq.~\eqref{homogeneous:asympo}, yields the asymptotic solution of~\citet{PocivavsekDellsy08}.

\subsection{Critical confinement for buckling}

Having determined the critical compressive force required to obtain buckling, $\tau_0(L)$, we now  determine the properties of the compressed sheet around this buckling threshold. The non--zero lower bound for the compression reveals that the sheet is subject to a uniform compression prior to buckling. We consider a sheet that is confined to a domain of size $2a$ slightly shorter than its undeformed length ($L$) such that its associated compression is $\tau_{\text{a}}<\tau_0(L)$. (Here, the subscript ``a'' is used to distinguish the applied compression needed to confine the sheet to a length $2a$ from the values required for buckling, i.e.~the eigenvalues of the mechanical equilibrium from Eq.~\eqref{homogeneous:ode0}.) Since $\tau_{\text{a}}<\tau_0(L)$ the sheet is stable in its flat configuration, i.e. $w(x)=0$, and is uniformly strained in the $x$-direction.  The horizontal strain $\epsilon_{xx}$ may be linked to the horizontal displacement $u(x)$ and also to the compressive stress $\tau_{\text{a}}$ via Hooke's law, giving
\be
	\frac{\partial u}{\partial x}=\epsilon_{xx}=-\frac{\tau_{\text{a}}}{\cal S},
\label{homogeneous:strain}
\ee which, upon integration from $-L/2$ to $L/2$ gives
\be
a=\frac{L}{2}\left(1-\frac{\tau_{\text{a}}}{\cal S}\right), 
\ee 
so that the critical value of $a$ at which the sheet buckles is
\be
a_c=\frac{L}{2}\left(1-\frac{\tau_0(L)}{\cal S}\right).
\ee

The sheet can accommodate any imposed compressive stress $\tau_{\text{a}}<\tau_0(L)$, or equivalently any confinement $a>a_c$, through in-plane compression. Below that length, the sheet buckles. As further confinement is imposed ($a<a_c$) the sheet's stress remains at the eigenvalue $\tau_0(L)$ and it tends to accommodate this further confinement by bending out of plane, rather than compressing in plane. We now consider how this amplitude grows just beyond the onset of buckling, i.e.~for $a<a_c$.

\subsection{Amplitudes\label{amplitudes}}

Having determined $\tau_0$ by solving Eqs \eqref{homogeneous:disprels}, the solution for the shape of the buckled sheet is given by Eqs. \eqref{homogeneous:sols} up to the multiplicative constants $A_{\text{s}}$ and $A_{\text{a}}$. Once the sheet has buckled out of plane, the expression for the strain in Eq. \eqref{homogeneous:strain} is modified to $\epsilon_{xx}=\tfrac{\upd u}{\upd x}+\tfrac{1}{2}\left(\frac{\upd w_0}{\upd x}\right)^2$ (see \tb{Appendix \ref{App:GovEqn}}). Integrating this from $x=-L/2$ to $x=L/2$ 
 and using symmetry of $(w_0')^2$ about $x=0$, results in
\begin{equation}
\int_0^{L/2}(w_0)_{,x}^2\mathrm{~d}x =L\bigl(1-\tau_0/{\cal S}\bigr)-2a=2(a_c-a).
\label{homogeneous:intrel}
\end{equation}
Using $w_0(x)$ from  Eqs.~\eqref{homogeneous:sole} and \eqref{homogeneous:solo} we find
\begin{widetext}
\begin{subequations}
\begin{align}
	\int_0^{L/2}\bigl[(w_0^{(\text{s})})_{,x}\bigr]^2\mathrm{~d}x&=A_{\text{s}}^2\left[\tfrac{L}{4}\bigl( k_+^2+k_-^2\bigr)+\tfrac{L}{2}k_+^2\tan^2 k_+L/2+k_+\tan k_+L/2\right],\\
	\text{or}\qquad \int_0^{L/2}\bigl[(w_0^{(\text{a})})_{,x}\bigr]^2\mathrm{~d}x&=A_{\text{a}}^2\left[\tfrac{L}{4}\bigl( k_+^2+k_-^2\bigr)+\tfrac{L}{2}k_+^2\cot^2 k_+L/2+k_+\cot k_+L/2\right],
\end{align}
\label{homogeneous:ampls}
\end{subequations}
\end{widetext}
which can be substituted into Eq.~\eqref{homogeneous:intrel} to give $A_{\text{s}}$ and $A_{\text{a}}$. Since $k_{\pm}$ and $\tau_0$ are determined by the numerical solution of Eq.~\eqref{homogeneous:disprele} or Eq.~\eqref{homogeneous:disprelo}, the amplitudes depend only on the value of $L$. Therefore, for given values of $a$ and $L$, the shape of the sheet can be completely determined numerically, including the amplitude, up to a sign.

We classify the sheets according to their size: Short length sheets are shorter than the intrinsic length scale of the wrinkles, $\ell^*$, ($\tL<1$), medium length sheets are a few length scales in size [$\tL=O(1,10)$], and large length sheets are many length scales in size ($\tL\gg 1$).
This classification is relevant to the choice of the displacement parallel to the direction of confinement.

\subsection{Switching between symmetric and antisymmetric modes \label{finite-size}}

For large $L$, the asymptotic analysis of \S\ref{sec:Asymptotics} shows how the sheet deformation switches between antisymmetric and symmetric modes whenever $L$ increases by $\pi$. A similar oscillation is observed in the numerical results for the smallest roots of Eqs.~\eqref{homogeneous:disprels}, plotted in Fig.~\ref{finite-size:modes} (a) for medium length sheets, and Fig.~\ref{finite-size:modes} (b) for large length sheets. However, this period of $\pi$ is only attained asymptotically as $L\to\infty$: for finite $L$ the length between mode switches must be calculated separately, as shown in \tb{Appendix \ref{App:Crossings}}.

When the modes switch, or their compressions cross as $L$ varies, the two relations~\eqref{homogeneous:disprels} are satisfied by the same values of $\tau_0$ and $\tL$.
The crossing points that we call type I are located at
\begin{equation}
        \tL=\sqrt{4l^2-1}\equiv\tL_{l,1}^{(\text{I})}\,, \label{finite-size:crossing1}
\end{equation}
with $l\in \mathbb{N}^\star$; index $l$ labels the crossings --starting at $l=1$ for the crossing at the smallest $\tL$, which is of type I, and increasing with the length of the sheet. Similarly, for what we call type II crossings we have
\begin{align}
        \tL=2\sqrt{l(l+1)}\equiv\tL_{l,1}^{(\text{II})}\,. \label{finite-size:crossing2}
\end{align}
The index $l$ also explicitly determines the corresponding compressions~\footnote{At the crossing points, the compressions are $\tau_0=\frac{2l+2j-1}{2l-1}+ \frac{2l-1}{2l+2j-1}$ and $\tau_0=\frac{l}{l+j}+ \frac{l+j}{l}$ respectively for type I and type II crossings. See \tb{Appendix \ref{App:Crossings}}.}.
The second index in Eqs.~\eqref{finite-size:crossing1} and~\eqref{finite-size:crossing2} evaluates to 1 because here we are interested in the first pair of compressions. The crossings for higher modes are included in \tb{Appendix \ref{App:Crossings}}.

For short length sheets ($\tilde L <1$) the compressions are inversely proportional to $\tilde L^2$ (see Fig.~\ref{finite-size:modes}a). This power--law behaviour can be understood by dimensional analysis; as the size of the confined sheet decreases it becomes the smallest length scale in the problem and hence $\tau^*_0\sim B^*_0/(L^*)^{2}$. However, a fundamental assumption for the floating beam equation \eqref{buckle:ode} is that the size of the sheet is larger than its thickness, and hence we will not explore the limit in which this condition is violated.

\section{Buckling of inhomogeneous sheets\label{sec:inhomcomposite}}

The results of the homogeneous case show that the amplitude of the buckling mode at the onset of buckling varies spatially, with a  maximum close to the centre of the sheet. One natural  question is: Can we control this amplitude variation, for example by introducing a stiffness gradient? To make this idea more concrete, we imagine adding liquid inclusions to our soft host such that there is a stiffness gradient within the beam, and solve the resulting problem numerically.

\subsection{Theoretical setting\label{sec:inhomcomposite-theory}}

\begin{figure}
	\includegraphics[width=\columnwidth]{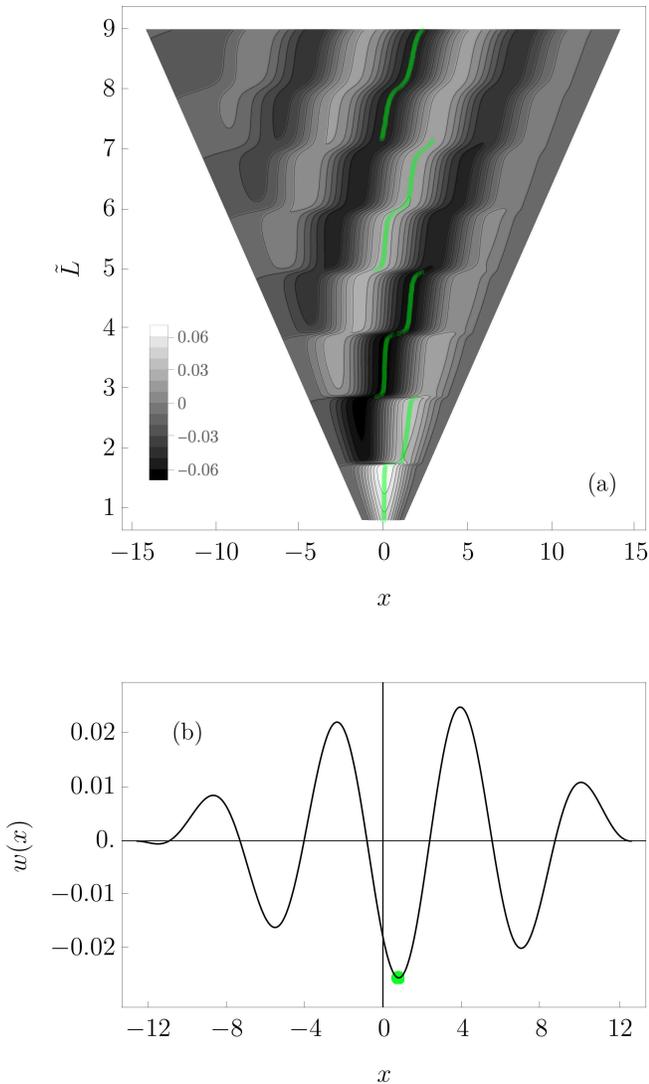}
	\caption{(a) Contour plots for $w(x)$ with varying $\tL$ and the largest vertical displacement (highlighted in green), for a medium length stiffened sheet. Here, $\phiend=0.1,\,\gamma\prime=10$.
	The amplitudes are computed assuming the condition $(a_c-a)=10^{-3}\ll1$, that ensures that the sheet bends close to the onset of buckling, imposed in Eq.~\eqref{homogeneous:intrel}. (b) Buckling profile in a sheet with $\tL=8$.}
\label{soft-res:profiles}
\end{figure}

Imposing a gradient of the liquid volume fraction parallel to the direction of confinement, $\phi(x)$, means that the elastic coefficients also vary spatially, i.e.~$E=E(x)$ and $\nu=\nu(x)$.  Therefore, we have a variable bending stiffness, viz.,
\begin{equation}
B^*(x^*)=\frac{E^*(x^*)\,{h^*}^3}{12[1-\nu(x^*)^2]}.\label{mech:bending_inhom}
\end{equation}Note that while $B^*$ may vary due to changes in the thickness $h^*$, we do not consider this possibility here, which would modify the isostatic floating requirement as noted above.

We anticipate that a gradient in $\phi$ will break the symmetries about $x=0$ exhibited by homogeneous sheets. Naively, one might expect the portion of the buckled sheet with the largest amplitude to reside where the modulus is smallest, since bending is easiest here, but  this poses further questions, such as where should such a sheet will experience maximum bending stress, and hence be most likely to fracture?

We consider for the numerical analysis a sheet that is embedded with stiffening and incompressible liquid inclusions ($\gamma\prime=10$, with $\gamma\prime\equiv l^*/R^*$ as per \tb{Appendix \ref{AppA}}; $R^*$ denotes the size of the inclusions), which are linearly distributed parallel to the direction of confinement:
 \begin{equation}
	\phi(x)=\phi_{\text{end}}\left(\frac{1}{2}-\frac{x}{L}\right),
 \label{buckle:phi}
\end{equation}
where $\phiend:=\phi(x=-L/2)$ is the concentration of inclusions at the left-hand end of the sheet.
 The scaled bending stiffness $B(x)$ equals the dimensionless effective Young's modulus for stiffened soft composites given by~\citet{StyleBoltyanskiy15,StyleWettlaufer15}:
\begin{equation}
	E^*[\phi(x),\gamma\prime]=E_{0}^*\,\frac{1+\frac{5}{2}\gamma\prime}{\frac{5}{2}\gamma\prime[1-\phi(x)]+\left[1+\frac{5}{3}\phi(x)\right]},\label{composite:Ec_DT}
\end{equation}
where $\phi(x)$ is given by Eq.~\eqref{buckle:phi}. The control parameters here are 
$\phiend$, and the size of the sheet, $L$.

\subsection{Spatial variation of amplitude\label{sec:inhomcomposite-amplitude}}
For medium length sheets [$\tL=O(1,10)$], we numerically solve the equation of mechanical equilibrium~\eqref{buckle:ode} using the {\it Chebfun} package~\cite{Chebfun} to obtain the buckling profiles corresponding to the smallest compression. The contour plot of $w(x)$ for different lengths $\tL$ in Fig.~\ref{soft-res:profiles} (a) shows that the maximal wrinkle amplitude (the trajectory in green) is shifted to the softer side of the sheet ($x>0$).
Indeed, the buckling profile at $\tL=8$, in Fig.~\ref{soft-res:profiles} (b), is no longer symmetric, unlike the modes of the homogeneous sheet ($\tL=8$) in Fig.~\ref{asymptotics:modes}.
This intuitive feature is also observed in the buckling profiles of an inhomogeneous column~\cite{SuneWettlaufer21}. We also see that the wrinkle wavelength is not noticeably altered by the change in elastic modulus. To understand this, we note that our asymptotic solution for large homogeneous sheets, leading to \eqref{homogeneous:asympe}--\eqref{homogeneous:asympo}, shows that the (short) wrinkle wavelength depends on the sum $k_++k_-$, while the large wavelength of amplitude modulations is controlled by the difference $k_+-k_-$; as such the fine scale wrinkle wavelength is affected only at higher order in spatial variation of properties than the amplitude modulation itself, as is observed. (To plot the function $w(x)$, we use the integral constraint Eq.~\eqref{homogeneous:intrel}, with $(a_c-a)=10^{-3}\ll1$; this is close to the onset of buckling where our theory is valid.)

Despite the lack of symmetry, the buckling profiles of a medium length inhomogeneous sheet are similar to the fundamental symmetric/antisymmetric modes characteristic of the homogeneous sheets. We project the numerical solutions for the inhomogeneous sheet, $w(x)$, onto the modes of the homogeneous sheet (Eqs.~\ref{homogeneous:sols}), i.e. the basis of eigenfunctions of the eigenvalue problem given by Eq.~\eqref{homogeneous:ode0} \footnote{Integration by parts shows that the operators $\left(\frac{\mathrm{d}^4}{\mathrm{d}x^4}+1\right)$ and $\left(-\frac{\mathrm{d}^2}{\mathrm{d}x^2}\right)$ are Hermitian, the latter is also positive definite, and hence $\{\left(\frac{\mathrm{d}^4}{\mathrm{d}x^4}+1\right),\left(-\frac{\mathrm{d}^2}{\mathrm{d}x^2}\right)\}$ is a Hermitian definite pencil. Therefore, the problem is one of standard Hermitian eigen-theory and we can take advantage of the completeness of the known eigenfunctions, Eqs. \eqref{homogeneous:sole} and \eqref{homogeneous:solo}, of the homogeneous sheet problem, Eq.~\eqref{homogeneous:ode0}.}. More concretely, we expand the solution of Eq.~\eqref{buckle:ode} as the infinite linear combination
\begin{align}
	w(x)=\sum_{j=1}^{\infty} \left[\mathrm{a}_j^{\text{(s)}} w_0^{(\text{s},j)}(x)+\mathrm{a}_j^{\text{(a)}} w_0^{(\text{a},j)}(x)\right], 
	\label{soft-res:complete}
\end{align}
where we use the index $j$ to denote each pair of solutions --- symmetric (s) and antisymmetric (a). We redefine the eigenfunctions~\eqref{homogeneous:sole} and ~\eqref{homogeneous:solo} as
\begin{align}
	w_0^{(\text{s})}(x)&=\cos k_-\frac{L}{2}\,\cos k_+x-\cos k_-x\,\cos k_+\frac{L}{2},\label{soft-res:sole}\\
	\text{and}\quad w_0^{(\text{a})}(x)&=\sin k_-\frac{L}{2}\sin k_+x-\sin k_-x\sin k_+\frac{L}{2},
\label{soft-res:solo}
\end{align}
and choose to normalize to unity the coefficients in the linear combination Eq.~\eqref{soft-res:complete}.

\begin{figure}
	\includegraphics[width=\columnwidth]{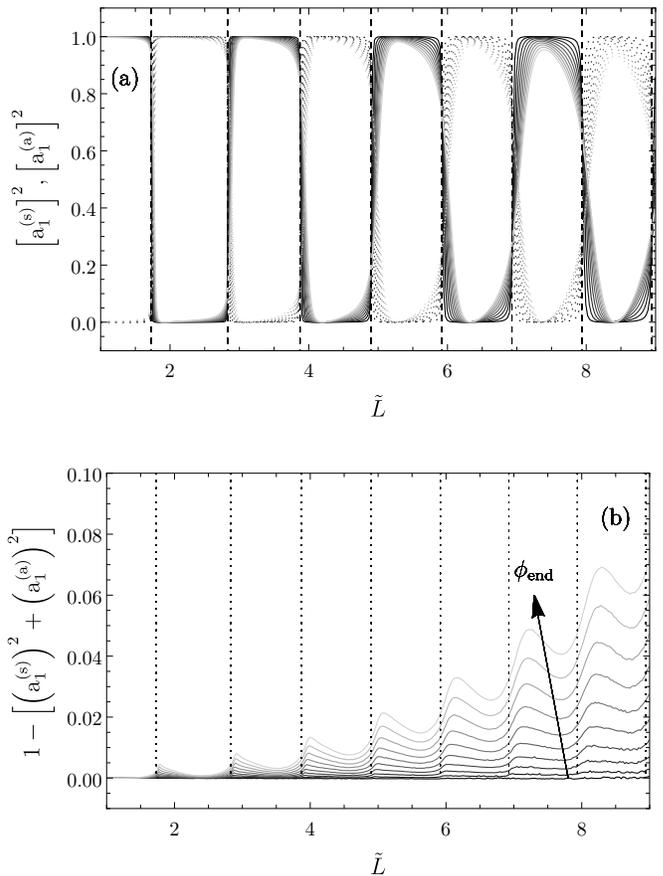}
	\caption{Smallest pair of squared coefficients in the expansion~\eqref{soft-res:complete} for a stiffened inhomogeneous sheet ($\gamma\prime=10$). (a) Solid curves denote $\Big[\mathrm{a}_1^{(\text{s})}\Big]^2$, and dotted curves $\Big[\mathrm{a}_1^{(\text{a})}\Big]^2$. (b) Residual that is not captured by the lowest symmetric and antisymmetric modes. Different shades of gray for trajectories corresponding to each value of $\phiend$, from $\phiend=0.02$ (black), to $\phiend=0.2$ (lightest gray), and equally spaced jumps for the curves in between (increasing in the direction of the arrow in (b)). The dashed vertical lines denote the crossing points as in Eqs.~\eqref{finite-size:crossing1} and~\eqref{finite-size:crossing2} ($j=1,l=\{1,2,3,4\}$).}
\label{soft-res:coeffs}
\end{figure}

The coefficients $\mathrm{a}_1^{\text{(s)}}$ and $\mathrm{a}_1^{\text{(a)}}$ are plotted in Fig.~\ref{soft-res:coeffs}a. The first pair of modes --- either the symmetric or the antisymmetric --- are dominant in the linear combination that describes $w(x)$. This dominant term changes in the vicinity of the crossing points (depicted by vertical dashed lines in Fig.~\ref{soft-res:coeffs}). 
However, for larger and more inhomogeneous sheets the residue that is not captured by the lowest symmetric and antisymmetric modes, $1-\Big[\Big(\mathrm{a}_1^{\text{(s)}}\Big)^2+\Big(\mathrm{a}_1^{\text{(a)}}\Big)^2\Big]$, plotted in Fig.~\ref{soft-res:coeffs}b, grows.
(Those cases with larger values of $\phiend$ are plotted in lighter gray in Figs.~\ref{soft-res:coeffs}.)
This indicates a reduction in the projection of $w(x)$ onto the first pair of modes.

\begin{figure}
	\includegraphics[width=\columnwidth]{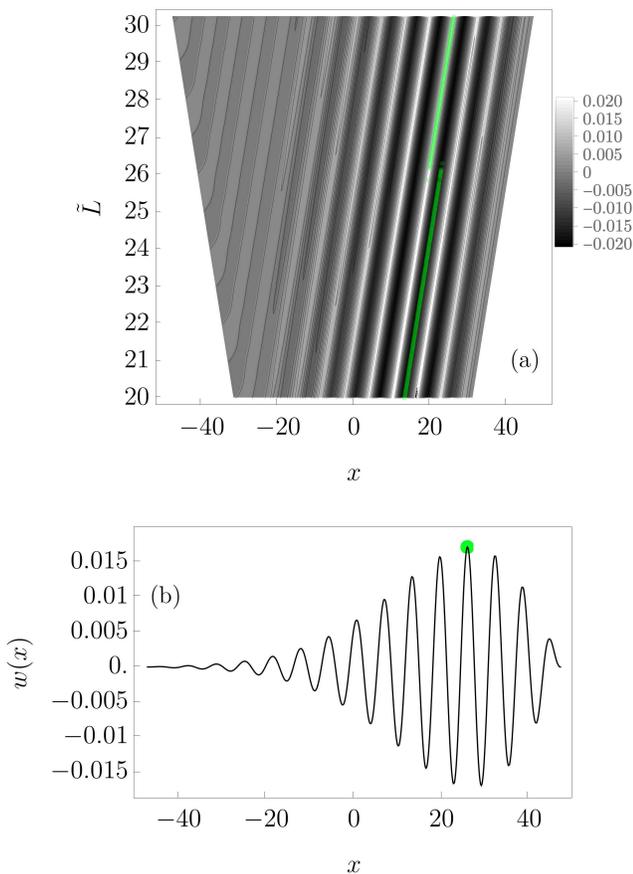}
	\caption{Displacement of the spatial variation in amplitude in the buckling profile $w(x)$ for a large length stiffened inhomogeneous sheet ($\phiend=0.1,\gamma\prime=10$) as $L$ varies. The sheet is stiffer at $x=-L/2$ and softer at $x=L/2$. (a) Contour plots for $w(x)$ for varying $\tL$ and the largest vertical displacement in green. (b) Buckling profile in a sheet with $\tL=30$.  $(a_c-a)=10^{-3}$ to compute the amplitudes from Eq.~\eqref{homogeneous:intrel}.}
\label{localization:profiles}
\end{figure}

Higher order modes are necessary to account for the buckling profile of longer and more inhomogeneous sheets. Thus the difference between the wrinkles in homogeneous and inhomogeneous sheets increases. This superposition of modes also explains the concentration of the bending energy towards the softer end of the sheet, as seen in Fig.~\ref{localization:profiles} for the long inhomogeneous sheet ~\cite{PocivavsekDellsy08,DiamantWitten11,Audoly11,OshriBrau15}.

\subsection{The failure of hard composites\label{sec:inhomcomposite-fracture}}
Our study of wrinkles thus far, leads to further questions.  For example,  when we consider brittle materials, such as ice, in the homogeneous case we expect failure (e.g.~fracture) to happen at the highest bending stress, i.e.~at the centre. If we introduce a stiffness gradient, we have shown above how the position of the maximum amplitude is displaced from the centre.  How does this position compare with that of the maximum bending stress? In other words, where is fracture more likely to occur in an inhomogeneous sheet?

Because ice is a hard material, we model the composite structure of water inclusions with the effective Young's modulus and the Poisson's ratio of~\citet{Eshelby57} as
\begin{align}
        E^*&\approx E_{0}^*\left(1-\left[\frac{3(1-\nu_{0})(1+13\nu_{0})}{(1+\nu_{0})(7-5\nu_{0})}\right]\phi\right),\label{composite:E_EP}~\text{and}\\
        \nu&\approx\nu_{0}+\left[\frac{12(1-\nu_{0})(-1+2\nu_{0})}{-7+5\nu_{0}}\right]\phi,\label{composite:Poisson_EP}
\end{align}
(see \tb{Appendix \ref{AppA}}). A model for fracture in ice floes is given in~\citet{Vella:2008}; when a crack is initiated, the stresses exceed a critical value $\sigma_{\text{m}}^*$~\footnote{For ice floes the yield strength, $\sigma^*_{\text{m}}$, is $1 – 3$ MPa for fresh ice~\cite{Hobbs10,Schulson99}, and $0.1 – 0.4$ MPa~\cite{WeeksAnderson58}, or 0.4 MPa~\cite{EvansUntersteiner71}, for sea ice.}.

For elastic sheets, stresses vary linearly through the thickness of the sheet, and so the maximum stress is achieved at the surface of the sheet.
This stress is related to the maximum bending moment per unit length acting on an element of the sheet~\cite[see pg. 5 of Ref.][]{Mansfield}:
\begin{align}
	\sigma_{\text{max}}^*=\frac{E^*(x^*)h^*}{2[1-\nu(x^*)^2]B^*(x^*)}|M^*(x^*)_{\text{max}}|\,,\label{localization:stress}
\end{align}
where the maximum bending moment per unit length is
\begin{multline}
	|M^*(x^*)_{\text{max}}|=\\
	\max\{B^*(x^*)|w^*_{,2x^*}(x^*)|:x^*\in[-L^*/2,L^*/2]\}\,,\label{localization:moment}
\end{multline}
where $w^*_{,2x^*}(x^*)$ is the curvature of the sheet in the small slope approximation (which is implicit in the derivation of the floating beam equation \eqref{buckle:ode}, \cite{Mansfield}). This maximum represents a trade--off between the bending stiffness, which peaks at the stiffer end of the sheet, and the curvature of the buckling profile, which is larger towards the softer side of the sheet.

\begin{figure}
	\includegraphics[width=\columnwidth]{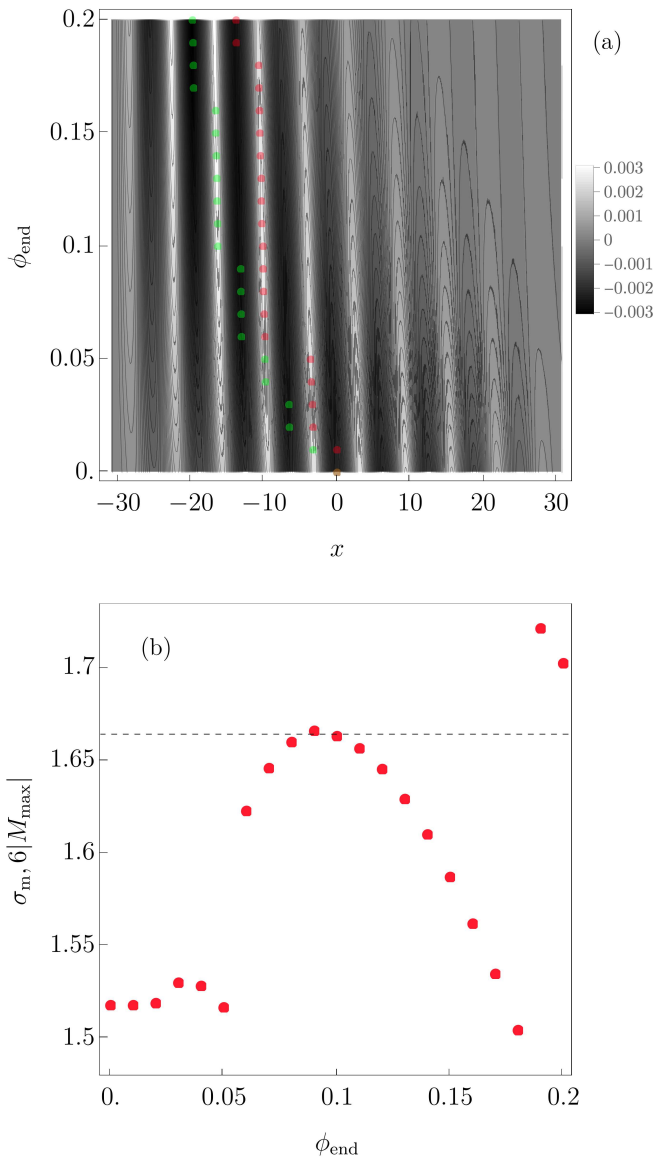}
	\caption{Bending profiles $w(x)$ and maximum bending stress of a thin, wide softened sheet ($\tL=19.6$) close to the onset of buckling ($(a_c-a)=10^{-5}$). (a)~Contour plots for $w(x)$ with varying volume fraction $\phiend$ are computed numerically~\cite{Chebfun}. The largest vertical displacement for every profile of a given $\phiend$ is marked by green dots. Red dots denote the position of the maximum bending stress. (Greyscale is used to show the value of $w(x)$, as indicated by the bar to the right.)  (b)~The maximum bending  stress (dots) and the rescaled version of the yield strength $\sigma^*_{\text{m}}=2.8~\text{MPa}$~\cite{Hobbs10,Schulson99} (horizontal dashed line).}
\label{localization:fracture}
\end{figure}

We rescale the bending moment per unit length by $B_{0}^* (\ell^*)^{-1}$, and the stress by $B_{0}^* (h^*)^{-2}(\ell^*)^{-1}$. Now, considering a sheet with constant thickness $h^*$ and varying Young's modulus, the dimensionless failure criterion is
\begin{align}
	\sigma_{\text{m}}\le 6 |M_{\text{max}}|\,.\label{localization:failure-scaled}
\end{align}

We plot in Fig.~\ref{localization:fracture} (a) the buckling profiles of a thin floating sheet with a decreasing volume fraction of softening inclusions (Eq.~\ref{buckle:phi} and elastic moduli given by Eqs.~\ref{composite:E_EP},~\ref{composite:Poisson_EP}) along the direction of confinement ($\tL=19.6$).
The vertical axis in the contour plot denotes an increasing volume fraction at $x=-L/2$, $\phiend$.
The dimensionless stretching-stiffness that models the behaviour of a fresh ice sheet, with $\rho^*_{\text{ice}}=0.9~\text{kg}\cdot\text{m}^{-3}$, $E_{0}^*=1~\text{GPa}$, $h^*=1~\text{mm}$ and $\nu_{0}=0.33$~\cite{Hobbs10,Schulson99}, floating on water, is a large value of $\mathcal{S}\approx 10^4$. To compute the amplitudes of the buckling profiles using Eq.~\eqref{homogeneous:intrel}, we assume $(a_c-a)=10^{-5}$.

In addition to the spatial variation of the wrinkle amplitude towards the softer end of the sheet, which is more prominent at larger values of the volume fraction $\phiend$, there is a deviation between the position of the largest deflection (green dots) and the coordinate of $|M_{\text{max}}|$ (red dots). To wit, the sheet's maximum bending stress is not at the wrinkle of largest amplitude, but at wrinkles on the stiffer side.

The maximum bending stress and the largest deflection in Fig.~\ref{localization:fracture}~(a) move stepwise towards the soft end as $\phiend$ increases, and the separation between their positions changes with $\phiend$.
Now, because the largest curvature occurs at the largest deflection, and $|M_{\text{max}}|$ is proportional to the curvature (see Eq.~\ref{localization:moment}) then $|M_{\text{max}}|$ is a discontinuous function of $\phiend$. In Fig.~\ref{localization:fracture}~(b), the discontinuities correspond to when the maximum bending stress jumps to the left and hence approaches the location of the largest deflection.

Finally, we note that the maximum bending stress, plotted in Fig.~\ref{localization:fracture}~(b), increases in inhomogeneous sheets.

\subsection{Critical compression \label{soft-res}}

The buckling profiles of an inhomogeneous sheet have no symmetry for any finite $L$. Thus, the switching between symmetric and antisymmetric modes observed in the homogeneous case as $L$ varies has no analogue in the inhomogeneous case. We recall that those sheet lengths at which the homogeneous modes switch correspond to degenerate values of the associated compressions, and hence these degeneracies should disappear in the inhomogeneous case. This is a well established result in quantum theory, where the jargon for degenerate eigenvalues is `level crossing', the simplest case of which was analyzed by~\citet{NeumannWigner1929}. From a mathematical perspective, the problem here is very similar:  the linear differential operator defining the mechanical equilibrium, Eq.~\eqref{buckle:ode}, can be parametrized as $\mathcal{A}(\phiend)=\mathcal{A}_{\text{H}}+\phiend\mathcal{A}_{\text{I}}$. When $\phiend=0$, this parametrization yields the homogeneous case, Eq.~\eqref{homogeneous:ode0}: $\mathcal{A}_{\text{H}}w_0+\tau_0(w_0)_{,2x}=0$. The problem of finding the eigenvalues of $\mathcal{A}(\phiend)$ by perturbation theory, which expresses each eigenvalue as a power series in $\phiend$, then shows the unification of the eigenvalues into a single multi--valued analytic function of $\phiend$, and the avoidance of crossing for $\phiend\neq 0$~\cite{BenderOrszag,Lax96}.

\begin{figure}
\input{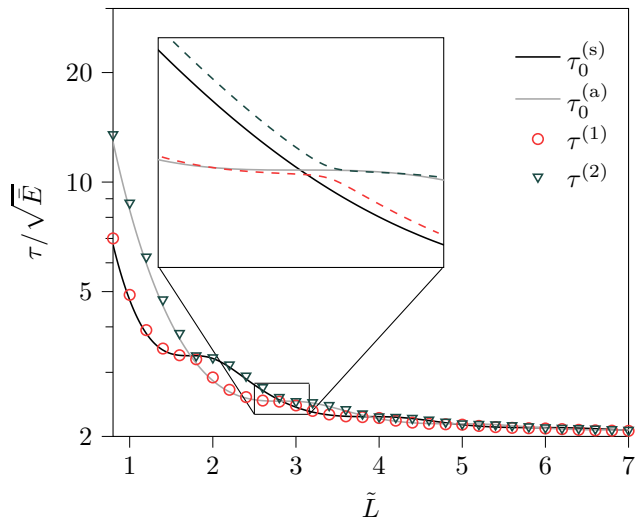}
	\caption{Critical compression of a stiffened inhomogeneous sheet ($\phiend=0.2,\,\gamma\prime=10$) as a function of the sheet size $\tL$, with numerical results using {\it Chebfun}~\cite{Chebfun} ($\tauu,\taud$). The quantities $\tau_0^{(\text{s})}$, $\tau_0^{(\text{a})}$ denote the smallest pair of compressions of the corresponding homogeneous sheet ($\phiend=0$). Here, the critical compression $\tau$ has been rescaled by the square root of the mean Young's modulus, $\tau/\sqrt{\bar{E}}$, to facilitate comparison with the results from the homoegenous case (solid curves).} 
	\label{soft-res:tau}
\end{figure}

We examine in Fig.~\ref{soft-res:tau} the two smallest compressions of inhomogeneous sheets as the sheet size varies. We use the rescaling $\tau/\sqrt{\bar{E}}$ ($\bar{E}$ denotes the spatial average of Eq.~\eqref{composite:Ec_DT}) in the inhomogeneous case and include for comparison the smallest pair of critical buckling compressions from the homogeneous sheet. We note that this scaling collapses the inhomogeneous case onto the homogeneous case. However, by blowing up the region around $\tL=\tL_{1,1}^{\text{(II)}}$ (see the inset in Fig.~\ref{soft-res:tau}, in which the inhomogeneous curves are computed every $\Delta\tL=0.032$), we see that the compressions of the inhomogeneous sheet do not cross. In other words, they are ordered from smaller to larger for any finite $\tL$. We thus denote the two smallest compressions $\tauu(\tilde{L})$ and $\taud(\tilde{L})$, with $\tauu(\tilde{L})<\taud(\tilde{L})$. Importantly, $\tauu(\tilde{L})$ is larger than the smallest compression in the first pair $(\tau_0^{(\text{s})}, \tau_0^{(\text{a})})$, and hence an inhomogeneous stiffness induces larger compressive loads; at most,  compressions that match the homogeneous counterpart are observed for certain values of $\tL$.

\begin{figure}
	\includegraphics[width=\columnwidth]{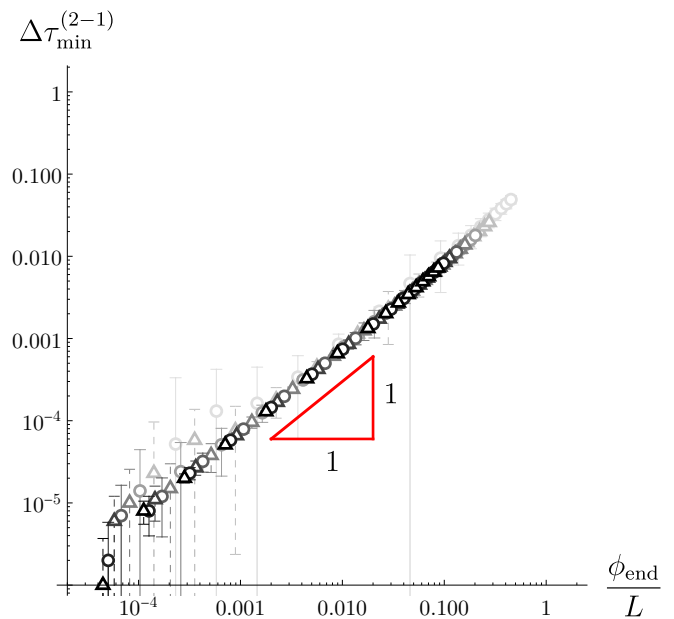}
	\caption{Minimal separation between the two smallest compressions 
	$\Delta\tau_{\text{min}}^{(2-1)}$ of a stiffened inhomogeneous sheet ($\gamma\prime=10$). Different symbols denote those minima corresponding to the crossings of type I (circles), and of type II (triangles) when $\phiend/L=0$. Shades of gray correspond to the crossing point index $l$: lighter gray for $l=1$, darkest gray for $l=4$. The red right triangle is one decade in length on both legs showing the linear relation 
	$\Delta\tau_{\text{min}}^{(2-1)}\propto\phiend/L$.}
\label{soft-res:dtmin}
\end{figure}

In a thought experiment in which we infinitesimally increase $\phiend$, starting at $\phiend=0$, i.e.~a homogeneous sheet, the intersecting compressions at $\tL$ given by Eqs.~\eqref{finite-size:crossing1} and~\eqref{finite-size:crossing2} split when $\phiend>0$. We measure this separation between consecutive compressions,  and we examine its growth with $\phiend$ close to the corresponding crossing points of the homogeneous sheet, in Fig.~\ref{soft-res:dtmin}. 

We fix $\phiend$ and then we find numerically the local minima of $\Delta\tau^{(2-1)}$ with respect to $\tL$, which we denote $\Delta\tau^{(2-1)}_{\text{min}}$. The values of $\tL$ that minimize $\Delta\tau^{(2-1)}$ start at the crossing points when $\phiend=0$, and increase with $\phiend$. The first eight results $\Delta\tau^{(2-1)}_{\text{min}}$ (corresponding to the first eight crossings in the homogeneous case) are plotted in Fig.~\ref{soft-res:dtmin} for different values of $\phiend/L$, where $L$ is the corresponding sheet size that minimizes $\Delta\tau^{(2-1)}$. All curves $\Delta\tau^{(2-1)}_{\text{min}}$ collapse into one. Therefore we conclude that $\Delta\tau^{(2-1)}_{\text{min}}$ grows linearly with $\phiend/L$, which is 
the gradient from Eq.~\eqref{buckle:phi}.

\section{Conclusions\label{sec:conclusions}}
We have examined the two--dimensional buckling and wrinkle patterns in floating homogeneous and inhomogeneous thin elastic sheets. With two control parameters, the size of the confined sheet and the gradient of bending stiffness, we quantified the wrinkled states using the F\"{o}ppl--von K\'{a}rm\'{a}n theory of thin sheets. A central test of the results is to vary the bending stiffness by varying the volume fraction of inclusions in the host solid. 

In homogeneous sheets, the only control parameter is the sheet size $L$. The buckling profile is determined by the smallest compressive load, and so we expect that the mode that will be observed in a buckling experiment corresponds to the smallest compression. However, for some particular sizes of the confined sheet, the same compression is associated with two different wrinkling modes. We gave asymptotic results for the shape of the sheet at  the onset of buckling, together with the critical loads and the critical sheet sizes for degeneracy in the limit of large sheets, $L\gg1$.

In contrast, this degeneracy is not observed in inhomogeneous sheets. Indeed, the otherwise crossing compressions of the homogeneous case split when a gradient of stiffness is applied parallel to the direction of confinement. The size of this splitting grows linearly with the magnitude of the gradient of the volume fraction at all the crossing points. Importantly, the wrinkled states of confined inhomogeneous sheets depend sensitively on their size. While medium length sheets buckle very much like their homogeneous counterparts, the wrinkled states in large length sheets are a superposition of many modes. This feature of large length sheets allows for the bending energy to be spatially concentrated, which is crucial in establishing a failure criterion, with particular relevance in glaciology.

Finally, the results presented here for floating sheets are also relevant for sheets on a linear soft elastic foundation,~\cite[see e.g.~Ref.][]{HuntWadee93}. A more complex behaviour is expected in the more general case of a linear elastic foundation, whose response is expected to be geometrically nonlinear, in which localization of buckling occurs,~\cite[see e.g., Ref.][]{Potier-Ferry1983,Brau2011}. In addition to the range of applications of interest, from soft composites of biological relevance \cite{GorielyBook} to hard composites of engineering or geophysical importance, a thorough mathematical analysis, rather than the numerical study given here, may provide additional insights. 

\begin{acknowledgements}
We thank A. Souslov and D. Mitra for helpful comments. MS, AFB and JSW acknowledge the support of Swedish Research Council Grant No. 638-2013-9243. CA acknowledges the support of the Swedish Research Council Grant No. 2018-04290. Nordita is partially supported by Nordforsk.
\end{acknowledgements}

\appendix
\begin{widetext}

\section{Details of theoretical formulation \label{App:GovEqn}}

 In the main text, we gave the equation governing the out-of-plane displacement of the beam without a formal derivation. Here, we expand upon the derivation of this.  A key detail concerns the state of stress within the sheet --- to ensure that this stress satisfies in-plane equilibrium, $\nabla\cdot\mathbf{\sigma}^*=0$, we introduce a force function $\varphi^*$: the internal in--plane forces per unit length are obtained by double differentiation of $\varphi^*$ so that $\sigma_{xx}^*=\partial^2\varphi^*/\partial y^{*2}$, $\sigma_{yy}^*=\partial^2\varphi^*/\partial x^{*2}$ and $\sigma_{xy}^*=-\partial^2\varphi^*/\partial x^*\partial y^*$ ~\cite{Mansfield}.  Following the standard derivation of the plate equation, see~\cite{Mansfield}, but accounting for the possibility that $\nu=\nu(x^*)$, we find that normal displacements of the sheet satisfy:
\be
	\nabla^2(B^*(x^*)\nabla^2 w^*)-[B^*(x^*)\{1-\nu(x^*)\},w^*]+\rho^*g^*w^*=[w^*,\varphi^*],
        \label{mech:gov}
\ee where $B^*(x^*)$ is the bending stiffness (or flexural rigidity) of the sheet and the von K\'{a}rm\'{a}n operator is
\be
    [w^*,\varphi^*]\equiv\pdTwo{w^*}{{x^*}}\pdTwo{\varphi^*}{{y^*}}-2\frac{\pd^2w^*}{\pd x^*\pd y^*}\frac{\pd^2\varphi^*}{\pd x^*\pd y^*}+\pdTwo{w^*}{{y^*}} \pdTwo{\varphi^*}{{x^*}}.
\label{mech:dieop}
\ee

For a midplane displacement with components $(u^*,v^*,w^*)$, the sheet's in--plane strains, $\epsilon_{ij}$, are given by
\begin{align}
        \epsilon_{x^*x^*}&=(u^*)_{,x^*}+\tfrac{1}{2}(w^*)_{,x^*}^2,\quad\epsilon_{y^*y^*}=(v^*)_{,y^*}+\tfrac{1}{2}(w^*)_{,y^*}^2,\nonumber \\
        \textrm{and}& \quad \epsilon_{x^*y^*}=\tfrac{1}{2}[(u^*)_{,y^*}+(v^*)_{,x^*}+(w^*)_{,x^*}(w^*)_{,y^*}],
\label{mech:strains}
\end{align} 
where, for example, $(w^*)_{,x^*}$ denotes the partial derivative of $w^*$ with respect to $x^*$. Note that the displacements $u^*$ and $v^*$ may be eliminated from these relationships by cross-differentiation,~\cite[see pg. 13 of Ref.~][]{Mansfield}. Relating these derivatives of strains to the in-plane forces, and hence to the derivatives of the force function $\varphi^*$, one finds the compatibility equation
\be
        \nabla^2\left(\frac{1}{E^*(x^*)}\nabla^2\varphi^*\right)-\left[\frac{1+\nu(x^*)}{E^*(x^*)},\varphi^*\right]=-\frac{h^*}{2}[w^*,w^*],\label{mech:eq_comp}
\ee
which gives the stress in the plane of the sheet induced by the stretching of the sheet's mid-plane.

In the two-dimensional buckling problem shown schematically in Figure \ref{mech:schema}, there are no variations in the $y^*$ direction (i.e.~into the page). Therefore, the vertical displacement of the sheet, $w^*$, is independent of $y^*$ and equations \eqref{mech:gov} and \eqref{mech:eq_comp} simplify to the following system;
\begin{subequations}
\begin{align}
	\frac{\mathrm{d}^2}{\mathrm{d}x^{*2}}\Bigl[B^*(x^*) w^*_{,2x^*}\Bigr]+w^*&=\varphi^*_{,2y^*}w^*_{,2x^*},\label{buckle:fvk:1}\\
	\textrm{and} \quad \nabla^2\left(\frac{1}{E^*(x^*)}\nabla^2\varphi^*\right)&=0.
\label{buckle:fvk:2}
\end{align}
\label{buckle:fvk}
\end{subequations}
Now, because $w^*=w^*(x^*)$ then $\varphi^*_{,2y^*}=f(x^*)$, so that $\varphi^*_{,x^*y^*}=A(x^*)+y^*f'(x^*)$. However, $\varphi^*_{,x^*y^*}$ is just the traction exerted on the sheet in the $y^*$ direction. This traction is zero for compression purely in the $x^*$ direction, so we have $f'(x^*)=0$, and hence $f(x^*)$ is a constant. Since the compressive load at the boundary $x^*=L^*/2$
is in the $x^*$ direction, and has magnitude $\tau^*$
\begin{align}
        \varphi^*_{,2y^*}(x^*=L^*/2)=-\tau^*,
\end{align}
and so that $f(x^*)=-\tau^*$. Thus Eq.~\eqref{buckle:fvk:1} becomes
\begin{equation}
        \frac{\mathrm{d}^2}{\mathrm{d}x^{*2}}\left[B^*(x^*) w^*_{,2x^*}\right]+\tau^* w^*_{,2x^*}+w^*=0,
\label{buckle:ode}
\end{equation}
which is Euler's linearized elastica equation \cite[see e.g. \S 20 of Ref.~][]{LandauLifshitz86} with a lateral load due to the hydrostatic pressure in the liquid foundation and varying elastic properties along the axis of confinement.
The boundaries of the thin sheet at $x^*=\pm L^*/2$ are clamped so that $w^*_{,x^*}(\pm L^*/2)=w^*(\pm L^*/2)=0$.

\section{Effective stiffness of composite materials \label{AppA}}
The foundational Eshelby theory of solid composites~\cite{Eshelby57} describes the elastic behaviour of rigid composites with a dilute dispersion of noninteracting incompressible inclusions.
Stiff-matrix materials such as ice, glass, ceramics and steel have $E^*=O(\text{GPa})$ and $\nu\sim0.3$, and thus have subnanometric elastocapillary length.  Therefore, for typical inclusion sizes the effect of surface tension is negligible and we can use Eshelby theory~\cite{Eshelby57} to compute the effective elastic moduli of compression, $\kappa^*$, and rigidity, $\mu^*$, which are
\begin{align}
        \kappa^*&=\kappa_{0}^*\left\{1- \left[\frac{(\kappa_{1}^*-\kappa_{0}^*)}{(\kappa_{0}^*-\kappa_{1}^*)\alpha-\kappa_{0}^*}\right]\phi\right\},~\text{and}\nonumber\\
        \mu^*&=\mu_{0}^*\left\{1-\left[\frac{\mu_{1}^*-\mu_{0}^*}{(\mu_{0}^*-\mu_{1}^*)\beta-\mu_{0}^*}\right]\phi\right\},\label{composite:Eshelby_moduli}
\end{align}
where $\alpha\equiv(1+\nu_{0})/[3(1-\nu_{0})]$ and $\beta\equiv2(4-5\nu_{0})/[15(1-\nu_{0})]$. We denote the host matrix's elastic constants with subscript ``0", those corresponding to the inclusions with subscript ``1", and symbols with no subscript denote the solid composite. The volume fraction of inclusions is $\phi$.

The incompressible liquid inclusions have zero shear modulus, $\mu_{1}^*=0$, and infinite bulk modulus, $\kappa_{1}^*=\infty$ (due to incompressibility), so that the Young's modulus and Poisson's ratio of the composite are
\begin{align}
        E^*&\approx E_{0}^*\left(1-\left[\frac{3(1-\nu_{0})(1+13\nu_{0})}{(1+\nu_{0})(7-5\nu_{0})}\right]\phi\right),\label{composite:E_EP_AppA}~\text{and}\\
        \nu&\approx\nu_{0}+\left[\frac{12(1-\nu_{0})(-1+2\nu_{0})}{-7+5\nu_{0}}\right]\phi,\label{composite:Poisson_EP_AppA}
\end{align}
	where we have used Eqs.~\eqref{composite:Eshelby_moduli} and expanded to first order in $\phi$. For a stiff-matrix composite with $\nu_{0}=0.3$ with liquid inclusions, the composite Young's modulus (Poisson's ratio) is less than (greater than) the host matrix;  $E^*<E_{0}^*$ and $\nu>\nu_{0}$.

For soft composites, micron sized inclusions create non-negligible interfacial stresses with an effective Young's modulus given by
\begin{equation}
	E^*(\phi,\gamma\prime)=E_{0}^*\,\frac{1+\frac{5}{2}\gamma\prime}{\frac{5}{2}\gamma\prime(1-\phi)+\left(1+\frac{5}{3}\phi\right)},\label{composite:Ec_DT_AppA}
\end{equation}
where $\gamma\prime\equiv l^*/R^*$ captures the size regime where surface tension operates~\cite{StyleBoltyanskiy15,StyleWettlaufer15}.  Eq.~\eqref{composite:Ec_DT} assumes that the inclusion concentration is dilute and hence we refer to it as the dilute theory (DT).  However, we note that this approach is quantitatively accurate up to $\phi\approx0.2$ when $\gamma\prime>2/3$, which is in the stiffening regime where $E^*(\phi,\gamma\prime>2/3)/E_{0}^*>1$~\cite{MancarellaStyle16b}.
The constituents of the composite are incompressible and hence $\nu=1/2$ throughout.

In the softening regime, where $\gamma\prime<2/3$, we use the expression for the effective Young's modulus of~\citet{MancarellaStyle16} (MSW):
\begin{equation}
	E^*(\phi,\gamma\prime)=E_{0}^*\,\frac{2-2\phi+\gamma\prime(5+3\phi)}{2+(4/3)\,\phi+\gamma\prime(5-2\phi)}.
 \label{composite:Ec_MSW}
\end{equation}

\section{The eigenvalue problem for a homogeneous sheet \label{AppB}}

Here we solve the linear fourth-order differential equation for the vertical displacement $w_0$, 
\be
\frac{d^4w_0}{dx^4}+\tau_0\ \frac{d^2w_0}{dx^2}+w_0(x)=0, \label{Eq}
\ee
where the eigenvalue $\tau_0$ is determined by the requirement to have a non-trivial that satisfies the homogeneous boundary conditions
\be
w_0(\pm L/2)=0\qquad\text{and}\qquad \frac{dw_0}{dx}\bigg|_{\pm L/2}=0. \label{BC}
\ee
Since equation (\ref{Eq}) has constant coefficients, we seek solutions of the form $w_0(x)\propto e^{ikx}$, yielding 
\be
k^4-\tau_0\ k^2+1 = 0, \label{quartic}
\ee
and the solutions for $k^2$ are 
\be
k_\pm^2 = \frac{\tau_0\pm\sqrt{\tau_0^2-4}}{2}\ \label{ksquare}
\ee or
\be
k_\pm=\frac{1}{2}\left(\sqrt{\tau_0+2}\pm\sqrt{\tau_0-2}\right).
\ee

Thus for the wavenumber $k$ to be real, we need $\tau_0\ge2$. The eigenfunctions $w_0(x)$ are linear combinations of $e^{\pm i k_+x}$ and $e^{\pm i k_-x}$, but we are interested in real solutions. Thus, we write the general solution of (\ref{Eq}) as 
\be
w_0(x)= C_1\ \cos(k_+x) +C_2\ \cos(k_-x) +C_3\ \sin(k_+x) +C_4\ \sin(k_-x), \label{eigenfunction}
\ee
where the $k_\pm$ are the positive roots of equation (\ref{ksquare}) and the $C_i$ $(i = 1, 2, 3, 4)$ are real constants. The problem is linear and has homogeneous boundary conditions, so those constants can only be determined up to a multiplicative factor. Enforcing Eqs. (\ref{BC}), we obtain
\begin{align}
C_1\  \cos(k_+L/2)+  C_2\  \cos(k_-L/2)&=0\nn\\
C_1\ k_+ \sin(k_+L/2)+  C_2\  k_-\sin(k_-L/2)&=0\nn\\
C_3\  \sin(k_+L/2)+  C_4\  \sin(k_-L/2)&=0 \label{syst}\\
C_3\ k_+ \cos(k_+L/2)+  C_4\  k_-\cos(k_-L/2)&=0.\nn
\end{align}
The vanishing determinant of the system (\ref{syst}) can be factorized as
\be
\begin{vmatrix}
\cos(k_+L/2) & \cos(k_-L) \\  k_+ \sin(k_+L/2) &  k_-\sin(k_-L/2)
\end{vmatrix}
\begin{vmatrix}
\sin(k_+L/2) & \sin(k_-L)\\  k_+ \cos(k_+L/2) &  k_-\cos(k_-L/2)
\end{vmatrix} 
=0.
\ee
Hence, for a given $L$ we have two possible relations involving $k_\pm$ and $L$:
\begin{align}
k_+ \tan(k_+L/2)&=k_- \tan(k_-L/2) \label{Srel}\\
\text{or}\qquad k_+ \cot(k_+L/2)&=k_- \cot(k_-L/2), \label{Arel}
\end{align}
	where `or' means that either the relation (\ref{Srel}) or the relation (\ref{Arel}) is fulfilled, or both. Note that when only Eq. (\ref{Srel}) is fulfilled, we must set $C_3=C_4=0$ for the last two equations in (\ref{syst}) to be satisfied, and hence the resulting eigenfunction is even. Similarly, when only Eq. (\ref{Arel}) is fulfilled, we must set $C_1=C_2=0$ and the eigenfunction is odd.
	
Clearly when $\tau_0=2$, $k_\pm=1$ and the relations \eqref{Srel} and \eqref{Arel} are both satisfied simultaneously. However, the resulting solution is trivial, $w_0(x)=0$. In \tb{Appendix \ref{App:Crossings}} we ask for which values of $L$ both relations (\ref{Srel}) and (\ref{Arel}) are satisfied simultaneously with $\tau_0>2$ and show that there are certain values of $L$, the `crossing points', for which both odd and even solutions emerge with the same value of $\tau_0$. More generally, however, one of the relations \eqref{Srel} and \eqref{Arel} has a smallest value of $\tau_0>2$. We therefore expect that as the compressive stress $\tau_0$ is increased from $0$, the mode with with the smallest value of $\tau_0$ will be obtained; this emergent buckling mode will therefore be symmetric or antisymmetric depending on which of the relations \eqref{Srel} and \eqref{Arel} is solved by the smaller value of $\tau_0$. In \tb{Appendix \ref{App:Asymptotics}} we determine asymptotic expressions, valid for $L\gg1$, for the smallest $\tau_0>2$ that satisfies each of \eqref{Srel} and \eqref{Arel}; this allows us also to determine which is the smaller compression and hence which mode, symmetric or antisymmetric, should be expected at the onset of wrinkling.

We are generally interested in the dependence of the eigenvalue $\tau_0$ on the natural length of the sheet $L$. We note that we can rewrite the characteristic equation (\ref{quartic}) as
\be
\tau_0= k^2 + \frac{1}{k^2}\ , \label{tau}
\ee
which holds for $k$ being either $k_+$ or $k_-$. We also note, from the original quartic, that  $k_+^2k_-^2=1$ and hence, taking $k_\pm$ to be positive, we have 
\be
k_+k_-=1. \label{krel}
\ee

\section{ The buckling wavenumber for homogeneous sheets is real\label{App:RealWaveNo}}

In \S \ref{homogeneous} we assumed that the wavenumber $k$ observed in buckling is purely real. Here, we demonstrate this is the case by supposing instead that Eq.~\eqref{quartic} has complex roots. Since the tension $\tau_0$ is real and Eq. (\ref{quartic}) contains only even powers, there must be two complex conjugate pairs of solutions.  We may write one pair as $k_\pm=k_r\pm\i k_i$, with $k_r\ge0$, and hence the other pair will be $-k_\pm$.
We extend sine and cosine to the complex plane using analytic continuation, and thereby extend the boundary conditions detailed in \tb{Appendix \ref{AppB}} to obtain the counterparts of Eqs. \eqref{Srel} and \eqref{Arel} as
\begin{align}
(k_r+\i k_i)\tan(k_r+\i k_i)L/2&=(k_r-\i k_i)\tan(k_r-\i k_i)L/2 \label{Sgen}\\
\text{and}\qquad(k_r+\i k_i)\cot(k_r+\i k_i)L/2&=(k_r-\i k_i)\cot(k_r-\i k_i)L/2, \label{Agen}
\end{align}
where for both relations (\ref{Sgen}) and (\ref{Agen}) the right-hand side is the complex conjugate of the left-hand side, so the imaginary part of either side must be zero. This condition takes the form
\be
f(k_r L/2) = - g(k_i L/2)\qquad\text{and}\qquad f(k_r L/2) = g(k_i L/2), \label{appcond}
\ee
for Eqs. (\ref{Sgen}) and (\ref{Agen}), respectively, where
\be
f(x) \equiv \frac{x}{\sin(x)\cos(x)}\qquad\text{and}\qquad g(x) \equiv \frac{x}{\sinh(x)\cosh(x)}\ .
\ee
By plotting the functions $f$, $g$ and $-g$ one finds that their ranges do not overlap (though $f$ and $g$ have the same limit as $x$ tends to zero) and hence there is no solution of Eq.~\eqref{appcond}. Therefore, $k$ is real.

\section{Asymptotic solution for large sheet sizes, $L\gg1$\label{App:Asymptotics}}

Having shown that the roots of Eqs. \eqref{Srel} and \eqref{Arel} are necessarily real, we consider in this Appendix the behaviour of these roots for large sheets, $L\gg1$. Our starting point is the observation, from numerical simulations, that as $L\to\infty$ it appears that $\tau_0\to2$ for both symmetric and antisymmetric modes. We therefore let
\be
\tau_0=2+\epsilon,
\ee
with $\epsilon\ll1$. From \eqref{homogeneous:waveno} we then immediately have that
\be
k_\pm=1\pm \tfrac{1}{2}\epsilon^{1/2}+\frac{\epsilon}{8}+O(\epsilon^2).
\label{eqn:kpmAsy}
\ee

We consider first the case of symmetric modes, rewriting Eq. \eqref{Srel} as
\be
\frac{k_+}{k_-}-1=-\frac{2\sin\bigl[(k_+-k_-)L/2\bigr]}{\sin\bigl[(k_++k_-)L/2\bigr]+\sin\bigl[(k_+-k_-)L/2\bigr]},
\label{eqn:SymRelKs}
\ee which can then be written in terms of $\epsilon$ as
\be
\epsilon^{1/2}+\frac{\epsilon}{2}+O(\epsilon^{3/2})=-\frac{2\sin\bigl[\epsilon^{1/2}L/2\bigr]}{\sin\bigl[1+\tfrac{1}{8}\epsilon+O(\epsilon^{2})\bigr]L+\sin\bigl[\epsilon^{1/2}L/2\bigr]}.
\label{eqn:SymRelEps}
\ee 
The quantity $\epsilon^{1/2}L$ appears frequently, and so we let
\be
\epsilon=\frac{4\alpha^2}{L^2}
\ee for some $\alpha$ which is an $O(1)$ quantity to be determined. We find that \eqref{eqn:SymRelEps} becomes
\be
\frac{2\alpha}{L}+\frac{2\alpha^2}{L^2}+O(L^{-3})=-\frac{2\sin\alpha}{\sin\bigl[L+O(L^{-1})\bigr]+\sin\alpha}.
\label{eqn:SymRelAlpha}
\ee A non-trivial solution requires $\sin\alpha\ll1$, and hence that $\alpha\approx n\pi$ for some integer $n$. The smallest $\tau_0>2$ corresponds to the smallest positive $\alpha$ so that the relevant root is $\alpha\approx\pi$. A simple calculation of the correction $\alpha-\pi$ from Eq. \eqref{eqn:SymRelAlpha} then yields the asymptotic expression for the value of $\tau_0$ for the even mode, $\tau_0^{(\text{s})}$, that is given in Eq. \eqref{eqn:Tau0AsyEven} of the main text. Precisely the same calculation, with minor modifications of signs on the right hand side of Eqs. \eqref{eqn:SymRelKs}--\eqref{eqn:SymRelAlpha}, follows through for the asymmetric mode and yields Eq. \eqref{eqn:Tau0AsyOdd} for $\tau_0^{(\text{a})}$.

We note further that the asymptotic expressions for $\tau_0^{(a)}$ and $\tau_0^{(s)}$ agree to leading order in $L^{-1}$ when $\sin L=0$; therefore we expect the crossing points to occur at $L=n\pi$ with $n\gg1$ integer.

The asymptotic expressions for symmetric and antisymmetric mode shapes, given in Eq. \eqref{homogeneous:asymp} of the main text, follow from expanding Eqs.~\eqref{homogeneous:sols} to leading order in $L^{-1}$.

\section{Crossings of symmetric and antisymmetric modes \label{App:Crossings}}

Here, we determine expressions for the values of $L$ for which both Eqs.~(\ref{Srel}) and (\ref{Arel}) are satisfied simultaneously; this corresponds to the sheet lengths for which a symmetric mode and an antisymmetric mode have the same eigenvalue. We refer to such a point as a `crossing'.

We multiply equations (\ref{Srel}) and (\ref{Arel}) to obtain $k_+^2=k_-^2$ and hence $k_+=k_-=1$. However, this implies $\tau_0=2$, which, as we have already seen, corresponds to the trivial solution. Instead, we must have that either $\tan k_+ L/2=\tan k_-L/2=0$ or $\cot k_+ L/2=\cot k_-L/2=0$, thereby allowing simultaneous solutions of Eqs. \eqref{Srel} and \eqref{Arel} with $k_+\neq k_-$.  Thus, there are two families of non--trivial common solutions to Eqs. (\ref{Srel}) and (\ref{Arel}):
\begin{enumerate}
\item $k_+L = \pi+2m\ \pi$ and $k_-L= \pi+2n\ \pi$, with $m,n\in \mathbb{N}$. Eq.~\eqref{krel} then implies that
\be
		L^{(\text{I})}=\pi\ \sqrt{(2m+1)(2n+1)}\  . \label{crossing1}
\ee
We refer to the eigenvalues $\tau_0$ at these crossing points as type I; they are given by (see Eq. \ref{tau})
\be
		\tau_0^{(\text{I})}=\frac{2m+1}{2n+1}+ \frac{2n+1}{2m+1}\ .
\ee
\item $k_+L = 2\tilde{m}\ \pi$ and $k_-L = 2\tilde{n}\ \pi$, with $\tilde{m},\tilde{n}\in \mathbb{N}^\star$ (the set of non-zero natural numbers). Eq.~\eqref{krel} implies that
\be
		L^{(\text{II})}=2\pi\ \sqrt{\tilde{m}\tilde{n}}\ . \label{crossing2}
\ee
We refer to the eigenvalues $\tau_0$ at these crossing points as type II; they are given by
\be
		\tau_0^{(\text{II})}=\frac{\tilde{m}}{\tilde{n}}+ \frac{\tilde{n}}{\tilde{m}}\ .
\ee
\end{enumerate}

Each symmetric and antisymmetric mode comes in a pair, which we label with the index $j=m-n\in\mathbb{N}^\star$ such that $j=1$ corresponds to the pair with the smallest eigenvalues. (Note that since $k_+>k_-$, $m>n$.) We find that the crossings within the pair $j$ occur for
\be
(m_j,n_j)= (l-1, l-1+j)\qquad\text{and}\qquad (\tilde{m}_j,\tilde{n}_j)= (l, l+j),\qquad  l\in\mathbb{N}^\star .
\ee
The index $l$ labels the crossings within a given pair, such that $l=1$ corresponds to the smallest size for which a crossing of either type occurs. Hence, we obtain two sets of crossing points given by
\be
L^{(\text{I})}_{l,j} =2\pi \sqrt{(l-1/2)(l+j-1/2)}\qquad\text{and}\qquad L^{(\text{II})}_{l,j} =2\pi\ \sqrt{l(l+j)}\ ,\qquad  l,j\in\mathbb{N}^\star. \label{matrices}
\ee
The corresponding eigenvalues are
\be
\bigr(\tau_0\bigl)^{(\text{I})}_{l,j} =\frac{2l+2j-1}{2l-1}+ \frac{2l-1}{2l+2j-1}\qquad\text{and}\qquad \bigr(\tau_0\bigl)^{(\text{II})}_{l,j} =\frac{l}{l+j}+ \frac{l+j}{l}\ .
\ee
For very small values of $L$, within a pair the symmetric mode always has the smallest eigenvalue. As $L$ increases, the first pair crosses at $L= \sqrt{3}\pi $ (type I), corresponding to $\tau_0 = 10/3$. The next crossing (type II) occurs at $L=2\sqrt{2}\pi$, corresponding to $\tau_0= 5/2$. Between those crossing points, the antisymmetric mode has the smallest eigenvalue. As $L$ increases this pattern repeats infinitely many times (see figure \ref{finite-size:modes}). Indeed, as $l\to\infty$, we note that:
\be
L^{(\text{I})}_{l,1}\sim 2\pi l,\quad L^{(\text{II})}_{l,1}\sim \pi (2l+1),
\ee reproducing the result of the asymptotic analysis for $L\gg1$ that followed on from Eqs. \eqref{eqn:Tau0AsyEven}--\eqref{eqn:Tau0AsyOdd}, namely that the system should switch between symmetric and asymmetric modes (and vice versa) each time $L$ increases by a multiple of  $\pi$.

\end{widetext}

\bibliographystyle{apsrev4-1}
\bibliography{wrinkle}

\end{document}